# The theoretical prediction of multifunctional Nb$_4$AC$_3$ (A=Al, Si, Ga, Ge): potential superconductors and unconventional optical properties

Teng Yue[a], Baochang Wang[b,‖], Weiwei Sun[c,*], Zhu Xiao-shuo[a], Feng Xiao-xue[a], Yan Mu-fu[d], Fu Yu-dong[a,*]

[a]School of Material Science and Chemical Engineering & Key Laboratory of Superlight Materials and Surface Technology, Ministry of Education,

Harbin Engineering University, Harbin 150001, China

[b]Department of Chemical Engineering, University College London, Torrington Place, London WC1E 7JE, United Kingdom

[c]Department of Physics and Astronomy, Material theory, Uppsala University, Box 516, SE-751 20 Uppsala, Sweden

[d]School of Material Science and Engineering, Harbin Institute of Technology, Harbin 150001, China

**Abstract**

Three new phases of ternary transitional metal carbides, namely MAX phases, Nb$_4$SiC$_3$, Nb$_4$GeC$_3$ and Nb$_4$GaC$_3$ were proposed by means of density functional theory, and they were all predicted to be mechanically stable. It was figured out that the α-type crystal structures of the hypothesized nanolaminates are energetically favored than β-type. Together with the known Nb$_4$AlC$_3$, the role of A group elements in Nb$_4$AC$_3$ (A= Al, Si, Ga, Ge) on various properties were systematically investigated. Not only bulk moduli, compressibility, shear moduli, and Young's moduli, but the approach to the ductility was discussed and raised as that the ductility can be achieved by forming Nb$_4$AC$_3$ containing more heavier $p$ electrons. All the studied Nb$_4$AC$_3$ phases exhibit the metallic nature that results from the Nb-4$d$ states being dominant at the Fermi level. The typical 4$d$-2$p$ hybridization between Nb-C leads to strong covalent bond and the contribution from the weaker 4$d$-3$p$(4$p$) hybridization between Nb-A are also identified. The $p$ electrons in A elements can effectively alter the covalency and the iconicity of the bonds that govern the compressibility, ductility and even superconductive properties. The extremely high population at the Fermi level in the density of states and the nesting effects in the Fermi surface shown in Nb$_4$SiC$_3$ both bring this material to a potential superconductor. The regular anisotropic behavior in the dielectric function is present but not in Nb$_4$SiC$_3$. Furthermore, Nb$_4$GeC$_3$ was also predicted to be a very promising candidate solar heating barrier material. The present work gives insight to the role of A elements on the mechanical, superconducting and optical properties of Nb$_4$AC$_3$ phases, and the obtained tendency by changing A elements will fertilize the desirable materials design of functional ceramics.

Correspondent authors: Fu Yu-dong (fuyudong@hrbeu.edu.cn) & Weiwei Sun (provels8467@gmail.com). The first two authors contributed equivalent work to this paper.

## 1. Introduction

Recently, much attention has been paid to the layered ternary compounds $M_{n+1}AX_n$ (n = 1,2,3), where M is an early transition metal, A is an A-group element and X is either carbon or nitrogen [1-6]. So far, a wide range of MAX phases (M can be $3d$, $4d$ or $5d$ transition metals), which have shown excellent electrical and thermal conductivity, damage-tolerant and resistant to oxidation and thermal shock and many other attractive physical and chemical properties by theoretical derivation, can be considered as potential candidates for high-temperature structural ceramics in wide areas of technological applications [7-15]. Upon to date, MAX phases have already served multifunctional materials in many applications fields. They are considered to replace graphite because of their promising attributes such as high strengths, moduli, wear and oxidation resistance and excellent thermal conductivities. MAX phases at this stage are of interest to work as polyfunctional ceramic thin films and coatings to enhance the performances of steels and alloys in various extreme conditions [16,17]. Their high stability withstanding cooling and heating render MAX phases adherent and protective heating elements. The low friction from room temperature to high temperature stimulate development of MAX based materials in tribological applications. Their unique combinations of metallic and ceramic properties make MAX phases and their composites meet the criteria to be used in the next generation of fossil energy power systems at a considerably reduced cost. Moreover, they have shown high mechanical fatigue/damage tolerance and good chemical compatibility with molten lead and sodium that are popular coolants for nuclear reactors. The high resistant to radiation damage (resulting form the mobility of dislocations in the basal plane) and the self-healing (dynamic recovery) at finite temperatures lead MAX phases to stand in the top list of potential materials in the nuclear industry. In addition to high-temperature applications mentioned above, there are electrical applications such as electrical contact deposition (developed by Impact Coatings, Sweden).

From the properties to the structures, MAX phases are mainly composed of $M_2AX$ (so called 211) and $M_3AX_2$ (so called 312) phases, with limited types of $M_4AX_3$ (so called 413) phases. The explorations and investigations on the new MAX phases provide many more basic scientific objects and in turn supply with possibilities for industrial applications, which has profound and constructive roles in long-term running. Up to now, only eight $M_4AX_3$ phases ($Nb_4AlC_3$, $Ti_4AlN_3$, $Ti_4GaC_3$, $Ti_4GeC_3$, $Ti_4SiC_3$, $Ta_4AlC_3$, $V_4AlC_3$ and $V_4AlC_{2.69}$) have been successfully synthesized experimentally [18-23]. Such few found $M_4AX_3$ phases is attributed to their high formation energies.

In reality, several strategies to conquer such obstacles have been used such as finding competing phases with lower formation energies by applying extreme conditions as well as finding new paths for preparations aided by theory. For instance, the synthesis of $Ti_4GeC_3$ was first intrigued by the prediction of metastable $Ti_4SiC_3$ in density functional theory, and the preparation of $Ti_4GeC_3$ (because of the neighboring element of Si) had been attempted subsequently in the form of epitaxial thin films using magnetron sputtering [22,24]. $Nb_4AlC_3$ that was synthesized by annealing $Nb_2AlC$ at 1700°C has shown excellent mechanical properties at high temperatures [25,26,27], which gives us a hint that the highly ordered MAX phases could show better performances than those in lower ordered ones. In analogous to the cases of $Ti_4SiC_3$ and $Ti_4GeC_3$, the road of finding $Nb_4SiC_3$ is encouraging to carry out theoretically.

It is interesting to see that most of superconductors of MAX phases tend to be Nb based $M_2AC$ ($Nb_2AlC$, $Nb_2SC$, $Nb_2AsC$, and $Nb_2SnC$) [28,29,30,31]. One question to be imposed is how about the highly ordered Nb based MAX phases? Such interesting phenomenon in 211 phases and the mystery of highly order phases have already stimulated the exploration of $Nb_{n+1}GeC_n$ phases (n=1,2,3), but the preparations of highly ordered $Nb_{n+1}GeC_n$ phases (n=2,3) unexpectedly failed [32]. We speculate that the conditions in experiments produced rather low energetic competing phases. The positive aspect is that these highly ordered phases were theoretically predicted to be dynamically stable [33], which allows us to take further step to investigations on the properties of $Nb_4GeC_3$ and $Nb_4GaC_3$. It is also implied that these phases are metastable and could be prepared experimentally in the future.

Holding the clues above, our explorations concentrate on the predictions of mechanical properties, electronic structure, and optical properties of $Nb_4AC_3$ (A=Al, Si, Ga, Ge). Since it was reported that the stronger M-X interaction determines many properties of MAX phases, the role of A elements has not been systematically studied and still unclear, even for 211 phases. This rarely mentioned issue of relatively weaker M-A interaction in comparison with the stronger M-X interaction is investigated. One of purposes of this work is to fill in the gap through the investigations of Nb4AC4 phases (A=Al, Si, Ga, Ge). Our results have showed the studied four structures are all mechanically stabile, which can be rather beneficial to synthesize desirable structural MAX phases and their alloys. Not only the role of A elements on the electronic structure was reported, the unique electronic structure in Nb4SiC3 that can be considered as the promising superconductors was unveiled. On the other hand, the optical properties of solids can help us study band structure, exactions, localized defects, lattice vibrations, *etc*. The dielectric function corrected by the Drude model together with the reflectivity of the studied structures was revealed. Significantly, very unusual behavior of electronic anisotropy was identified in $Nb_4SiC_3$, which is relevant to the electronic structure. In a nutshell, this study will contribute deep understandings to Nb based $M_4AX_3$ phases, and the significance of A elements is verified and should be accounted for the future studies of MAX phases. The predicted tendency to better performance on mechanical, onset of superconducting, and unconventional optical properties will further stimulate investigations on $M_4AX_3$ phases, and provide significances to the material design of the multifunctional ceramics.

## 2. Methodology

The main part of calculations were accomplished using the Cambridge Serial Total Energy Package (CASTEP) code [34] based on density functional theory (DFT) with the generalized gradient approximation(GGA) in the scheme of Perdew-Burke-Ernzerhof [35]. All computations were performed in the framework of Vanderbilt-type ultrasoft pseudopotential [36], and the valence electronic configurations were presented by $2s^22p^2$ states for C, $3d^{10}4s^24p^1$ states for Ga, $4s^24p^2$ states for Ge and $4s^24p^64d^45s^1$ states for Nb. The Broyden-Fletcher-Goldfarb-Shanno (BFGS) minimization scheme was used to minimize the total energy and internal forces [37]. Geometry optimization was completed using convergence thresholds of total energy within $5\times10^{-6}$eV, maximum force within $0.01$eV/Å$^{-1}$, maximum stress within 0.02 GPa and maximum displacement within $5\times10^{-4}$Å. The plane-wave cutoff energy was tested and the value of 450 eV is sufficient. For the sampling of integration over the Brillouin zone, the $10\times10\times2$ *k*-mesh was fixed according to the Monkhorst-Pack scheme [38]. With the value of the plane-wave cutoff energy and the numbers of k points, the convergence tests showed good convergence of the computed energies (the total energy deviation less than 0.001%). The stress–strain method were employed to compute elastic constants that can be used to derive many significant values to evaluate mechanical properties. For metallic systems, several exotic phenomena can be reflected by Fermi surface, and therefore in order to provide a more accurate description, full potential local orbital code (FPLO version 9.01.35) [39] was employed to compute density of states, Fermi surface, and the plasma frequency. The *k*-mesh points mapped to study the electronic structure in FPLO were doubled as the number used in CASTEP. Note that the other computational parameters in FPLO remained the same with those in CASTEP.

## 3. Results and discussion

### 3.1. The predicted structures

Most of $M_4AX_3$ phases such as $Nb_4AlC_3$ possess two types of structures: α and β forms whose stacking sequences are different along the out-of plane direction, and these two different types of structures of $Nb_4AlC_3$ are taken as an example to be shown in Fig.1. The atomic arrangement of α phase is presented by AB**A**BACB**C**BC, and that of β phase is presented by AB**A**BABAB**B**AB. This shift of the M and X layers may tailor properties of these materials, so that the energy difference between α and β phases of all the studied phases should be address first, and the lattice parameters as well as atom positions are also presented in Table1.

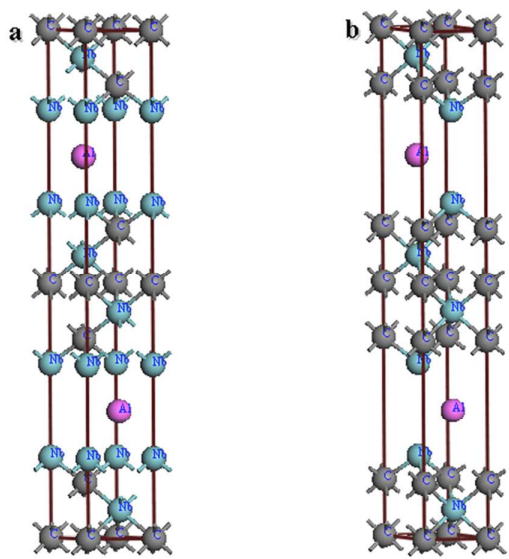

Fig. 1 Crystal structures of (a) α-$Nb_4AlC_3$ and (b) β-$Nb_4AlC_3$. The blue spheres present Nb atoms, pink for Al atoms, and grey for C atoms.

As Table 1 shows, both of β phases have higher energy of around 1eV than α phase, and it can be concluded that α phases for the studied materials are the stable crystal structure. The relative energy (ΔE) is calculated by E(α-$Nb_4AC_3$) – E(β-$Nb_4AC_3$). The formation energies ($E_{form}$) of these structures are calculated from the difference between their total energies and the sum of the isolated atomic energies of the pure constituents (fcc Al metal, bcc Nb metal, orthorhombic Ga (space group Cmca), fcc Ge as well as the diamond structure for C and Si). Furthermore, these computed negative formation energies shown in the last column in Table 1 demonstrates that α phases are energetically favored, which indicates that in experiments, they could be successfully synthesized through suitable competing phases. Based on the obtained results, α-$Nb_4AC_3$ (A=Ga, Ge) have relatively large negative formation energies than the other two structures suggesting that forming α-$Nb_4AC_3$ (A=Ga, Ge) is much more easier over that of α-$Nb_4SiC_3$.

Table 1.The optimized structural parameters, and the relative energies (ΔE) , formation energies with respect to the ground state of the corresponding elements of these four $Nb_4AC_3$ phases. Note that $z_1/z_2$ (Nb,C) indicate the variables of atomic coordinate of different types of Nb, C atoms.

| *Compound* | *c/a* | *$z_1/z_2$(Nb)* | *$z_3$(C)* | *$V_0$ ($Å^3$)* | *ΔE (eV)* | *$E_{form}$/f.u.(eV)* |
|---|---|---|---|---|---|---|
| α-$Nb_4GaC_3$ | 7.63 | 0.0555/0.1 | 0.1092 | 206.95 | 0 | -5.18 |
| β-$Nb_4GaC_3$ | 7.85 | —— | —— | 206.49 | 1.35 | —— |
| α-$Nb_4GeC_3$ | 7.42 | 0.0564/0.1 | 0.1105 | 205.49 | 0 | -5.39 |
| β-$Nb_4GeC_3$ | 7.56 | —— | —— | 204.27 | 1.01 | —— |
| α-$Nb_4AlC_3$ | 7.71($7.72^{[25]}$) | —— | —— | 204.59 | —— | -5.33 |
| α-$Nb_4SiC_3$ | 7.23 | 0.0579/0.1 | 0.1131 | 203.02 | 0 | -0.16 |
| β-$Nb_4SiC_3$ | 7.23 | 0.0579/0.1 | 0.1131 | 203.02 | 0.60 | —— |

In the α phase, Nb atoms occupy 4f (1/3, 2/3, $z_1$) and 4e (0, 0, $z_2$) sites, and C atoms occupy 2a (0, 0, 0) and 4f (2/3, 1/3, $z_3$) sites, as well as Al atoms occupying at 2c (1/3, 2/3, 1/4) site. As *c/a* value varies for each structure, α-$Nb_4GaC_3$ is found to be the most anisotropic structure with the highest *c/a* value of 7.85 among them. Note that the calculated *c/a* and volumes of $Nb_4AC_3$ (A=Ga, Ge) are within the range of the $Nb_4AC_3$ (A=Al, Si) group, which implies the reasonability of our prediction. The minor deviation from previous work in the case of $Nb_4AlC_3$ illustrates that a good agreement has been achieved. It is observed that the filling of one more *p* electron will bring out a smaller volume, e.g. $Nb_4AlC_3$ to $Nb_4SiC_3$, and $Nb_4GaC_3$ to $Nb_4GeC_3$. $Nb_4SiC_3$ is the most compact structure with the volume of 203.03 $Å^3$, and it is also the least anisotropic structure with a *c/a* value of 7.23. This can be understood in terms of atomic radius. Si has the smallest atomic radius, which can result in smallest lattice expansion. In contract, Ga has the largest atomic radius with largest volume (206.95 $Å^3$). As for the internal coordinates of Nb and C atoms, the variation of $z_1$, $z_2$ and $z_3$ sites are rather small, which supports the fact of the stability of this structure. It should be emphasized that in the following analysis, all results and discussion point to α phases.

In order to further characterize these structures in detail, the simulated X-ray diffraction patterns (XRD) of these four phases are plotted in Fig.2. The experimental diffraction pattern of $Nb_4AlC_3$ can be used as the reference [25], and our calculated data of $Nb_4AlC_3$ (blue curves in the bottom of Fig.2 a) are in the good agreement with the experimental data (green marks). It can be concluded that diffraction patterns of the studied phases are quite similar but with a few differences. The filling of one more $p$ electron in the same row (Fig.2 a-b, c-d) can push downwards the peaks above 50°, and the dispersive peaks between below 30° and above 45° do not show much change by this effect. The heavier $p$ electron can slightly alter the peak at 58° (Fig.2 a-c, b-d), which are split in the cases of $Nb_4GaC_3$ and $Nb_4GeC_3$. In addition, the new peak at 45° is only observed in the cases of $Nb_4AlC_3$ and $Nb_4GaC_3$. Normally, the shift of the positions of peaks can be attributed to lattice expansion or compression and thermal effects, and the emergence of new peaks at 30° and 45° can be regarded as the fingerprint of these structures. These features are supposed to be confirmed by experiments. From Fig. 2, it can be seen that the strongest peaks are situated at 38° and 40° that indicate (105) and (106) judging from the experimental data of $Nb_4AlC_3$ [25]. According to Bragg's law, the scanning angle (2θ) is inversely proportional to the inter-plane distance being a index of the volume for this layered materials. It is found that the 2θ=40° of $Nb_4SiC_3$ exhibits the largest value indicating the volume is likely to be the smallest, which is coincided with the values of volumes in Table 1.

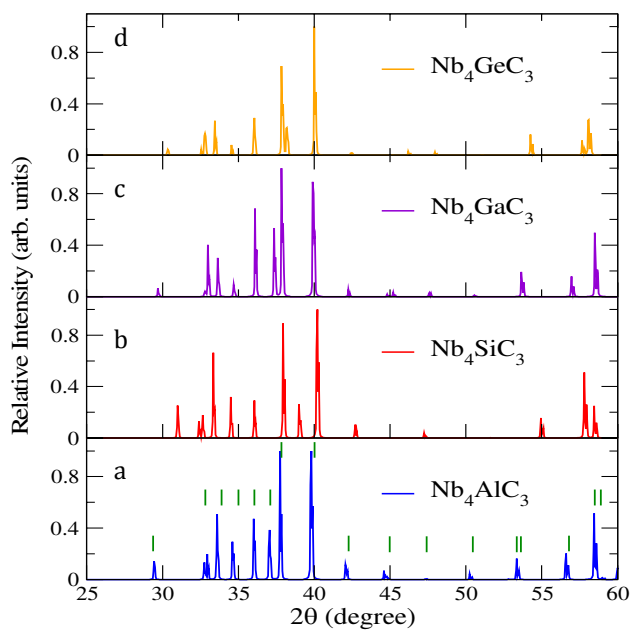

Fig. 2 X-ray diffraction patterns of α-$Nb_4AlC_3$ (in blue), α-$Nb_4SiC_3$ (in red), α-$Nb_4GaC_3$ (in violet) and α-$Nb_4GeC_3$ (in orage). The green vertical marks are the position of experimental Bragg peaks of $Nb_4AlC_3$ [25].

### 3.2. Stability and relevant mechanical properties

The elastic constants $C_{ij}$ are defined by means of a Taylor expansion of the total energy for the system, $E(V, \varepsilon)$, with respect to a small strain n of the lattice ($V$ is the volume) by applying a distortion to the lattice. The distortion matrix applied to the lattice vectors is written as:

$$\mathbf{e} = \begin{pmatrix} 1+\varepsilon_{xx} & \varepsilon_{xy} & \varepsilon_{xz} \\ \varepsilon_{yx} & 1+\varepsilon_{yy} & \varepsilon_{yz} \\ \varepsilon_{zx} & \varepsilon_{zy} & 1+\varepsilon_{zz} \end{pmatrix} \qquad \text{Equation (1)}$$

Normally, the Voigt notation replaces xx by 1, *yy* by 2, *zz* by 3, *xy* (and *yx*) by 6, *xz* (and *zx*) by 5, and finally *yz* (and *zy*) by 4 leading to the general $C_{11}, C_{22},$ etc. The obtained elastic constants can be applied to judge the mechanical stability. As the hexagonal crystals, the elastic constants of stable structures should meet the Born criteria [40]:

$$\begin{cases} C_{11} > 0, C_{44} > 0 \\ C_{11} - C_{12} > 0 \\ (C_{11} + C_{12})C_{33} - 2C_{13}^2 > 0 \end{cases} \qquad \text{Equation (2)}$$

It is clear that $Nb_4SiC_3$, $Nb_4GaC_3$ and $Nb_4GeC_3$ are all stable according to the above criteria based on the results in Table 2 shown below. Moreover, elastic constants can help us derive many other mechanical properties, such as bulk modulus $B$, shear modulus $G$, Young's modulus $E$ that were also calculated and shown in Table 2. The bulk modulus $B$ will be investigated firstly among these quantities. In table 2, it can be seen that the bulk modulus of $Nb_4GaC_3$ and $Nb_4GeC_3$ (200GPa and 250 GPa, respectively) are slightly greater than that of $Nb_4AlC_3$ (214 GPa) but less than that of $Nb_4SiC_3$ (235 GPa). It was reported that the high moduli of MAX phases are governed by the strong covalent bonding between M and X atoms [30–42]. It is interesting to unveil if the M-X interaction can be perturbed by M-A (details in the next section). From the changes of $B$ with A group elements (Al to Ga & Si to Ge), it can be inferred that the filling of one more *p* valence electron and the heavier *p* electron can harden the lattice and make the structure show stronger tolerance of compressibility. The similar trend is also observed in α-$Ti_4GaC_3$ and α-$Ti_4GeC_3$. To clarify the trend of bulk modulus, detailed analysis based on bond stiffness [43] has been employed and shown in Fig.3. In fact, the bond stiffness can be regarded as the slope of a quadratic curve obtained by $d/d_0 = C_0 + C_1 P + C_2 P^2$, where $d$ stands for the bond length, $d_0$ is the bond length at ambient conditions, and $P$ is the pressure. As the definition shows, at the ambient conditions, the bond stiffness at 0 GPa can be obtained by computing $1/C_1$. Because of the multiple bond types, several classes have been taken into account, but Nb (4f sites)-A bonds are ruled out owing to their long bond lengths (4 Å at ambient conditions). Therefore, the stiffness of Nb-C, Nb-A and the total stiffness are plotted in Fig. 3. The two components are compared to each other, and it is concluded that the stiffness of Nb-A is much lower than that of Nb-C, which can be explained from the electronic structure analysis in the next section. $Nb_4SiC_3$ has the highest bond stiffness, which corresponds to the largest bulk modulus shown in Table 2. Obviously, there is a deep drop in Nb-Ga, which refers to the weakest Nb-Ga bond among the four studied materials. The slight change in the total stiffness of $Nb_4AlC_3$, $Nb_4GaC_3$ and $Nb_4GeC_3$ shows a good agreement with the tendency of bulk moduli in Table 2. It is known that the covalent bond shows the strongest stiffness, followed by the ionic bond, then the metallic bond. Hence, the drastic drop of Nb-Ga stiffness can be due to the ionic feature between Nb-Ga (This will be discussed later in Fig.5 and Table 4). Overall, the tendency of the total bond stiffness with different A elements results in the increasing of bulk moduli, which is found in Ref. [43].

Table 2. Elastic constants $C_{ij}$, bulk moduli B, shear modulus G, Young's modulus E (all in GPa), and ratio G/ B

|  | $C_{11}$ | $C_{12}$ | $C_{13}$ | $C_{33}$ | $C_{44}$ | $C_{66}$ | B | G | E | **Ref.** |
|---|---|---|---|---|---|---|---|---|---|---|
| α-$Nb_4AlC_3$ | 428 | 109 | 123 | 355 | 168 | 159 | 213 | 156 | 374 | this work |
| α-$Nb_4AlC_3$ | 413 | 124 | 135 | 328 | 161 | 145 | 214 | 144 | 344 | [30] |
| α-$Nb_4SiC_3$ | 388 | 149 | 170 | 356 | 186 | 120 | 235 | 136 | 342 | This |
| α-$Nb_4GaC_3$ | 424 | 123 | 136 | 346 | 155 | 151 | 220 | 154 | 375 | this work |
| α-$Nb_4GeC_3$ | 404 | 149 | 153 | 319 | 161 | 128 | 225 | 132 | 332 | this work |
| α-$Ti_4GaC_3$ | 353 | 68 | 57 | 276 | 130 | 143 | 150 | 134 | 332 | this work |
| α-$Ti_4GeC_3$ | 355 | 74 | 80 | 321 | 138 | 141 | 167 | 137 | 327 | this work |
| $V_4AlC_3$ | 458 | 107 | 110 | 396 | 175 | 176 | 218 | 170 | 414 | [42] |

Compared with *B*, changes of G show an opposite behavior, thereby $Nb_4AlC_3$ has the highest shear modulus (156 GPa) and Nb$_4$SiC$_3$ is the smallest one (129 GPa). The shear modulus $G$ measures resistance to the shape change, and the Young's modulus is defined as the ratio between stress and strain. From the trend of bond stiffness in Fig. 3, there is no clear and direct connection with them. However, the tendency of $E$ with different A elements is very similar to that of $G$, which shows the opposite tendency to that of bulk moduli. As can be seen in Table 2, Nb$_4$AlC$_3$ is the stiffest one with a value of 374 GPa, followed sequentially by Nb$_4$GaC$_3$, Nb$_4$GeC$_3$, and the least one, Nb$_4$SiC$_3$ with value of 329 GPa. If the investigation on the relation of $E$ with $G$ and $B$, the Young's modulus is defined as $E = \dfrac{9BG}{3B+G}$. From the above relation, E is proportional to $G$ and inversely proportional to $G/B$. As it is disused above, the change of B is opposite to that of $G$, so that in the studied materials, the competition between $G$ and $G/B$ determines Young's moduli. As G and G/B are related to the ductility, it is natural to include some quantities describing plasticity to explain the tendencies of $G$ and $E$. From the view of structure, the strong anisotropic characteristic for MAX phases may make the stress-strain relation rather complicated, and further findings of new members in the family of Nb$_4$AC$_3$ will be also helpful to explain it. In a nutshell, all the studied Nb$_4$AC$_3$ show strong compressibility and ability to tolerate shape change, which are of significance to be used in several applications for example, hard coatings, ceramic composite matrix, etc.

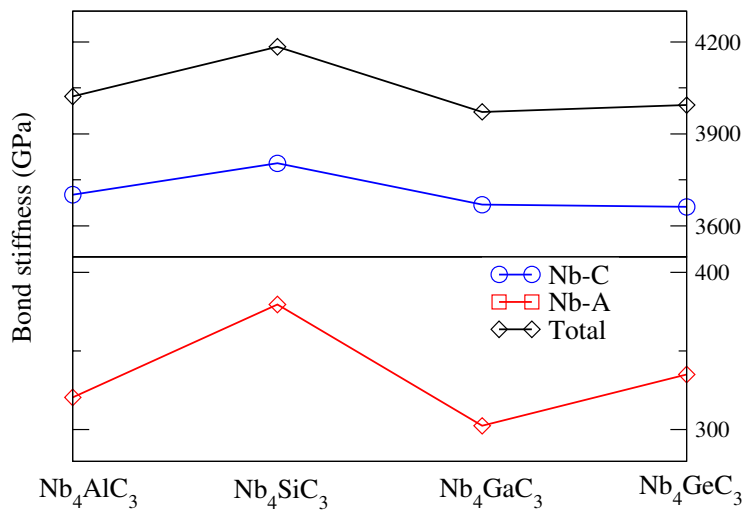

Fig. 3 The bond stiffness of Nb-C (blue circles), Nb-A (red squares) and Total (black diamonds) at 0 GPa in Nb$_4$AlC$_3$, Nb4SiC$_3$, Nb$_4$GaC$_3$ and Nb$_4$GeC$_3$.

It was reported that MAX phases are unusual to other ceramic materials, since they are machinable, like metals due to the M-A interaction. In this scenario, MAX phases bridge metals and ceramics together owing to the interstation of A element layers. It is therefore quite essential to estimate the brittleness or ductility for these materials. The *G*/*B* ratio and Cauchy pressure ($C_{12}$ – $C_{44}$), referring to $\Delta C$, are the two key parameters to assess ductility trends in conjunction with Pettifor [44] and Pugh [45] criteria. It is stated that the positive Cauchy pressure and *G*/*B* ratio below 0.5 indicate a ductile material. In Fig. 4, none of $M_4AX_3$ phases is ductile according to the mentioned criteria, but the $M_4AX_3$ phases containing Nb (4*d* M element) are less brittle than those containing 3*d* M elements (see the arrow). Among all the studied phases, $Nb_4GeC_3$ is the least brittle phase with *G*/*B* of 0.60. Compared to $Nb_4GeC_3$, $Nb_4GaC_3$ is more brittle, which is similar to the neighboring phases $Nb_4SiC_3$ and $Nb_4AlC_3$ in Fig 4. Therefore, the heavier A elements can increase ductility in $Nb_4AC_3$, and structures like $Nb_4SnC_3$ or $Nb_4AsC_3$ are likely to be ductile.

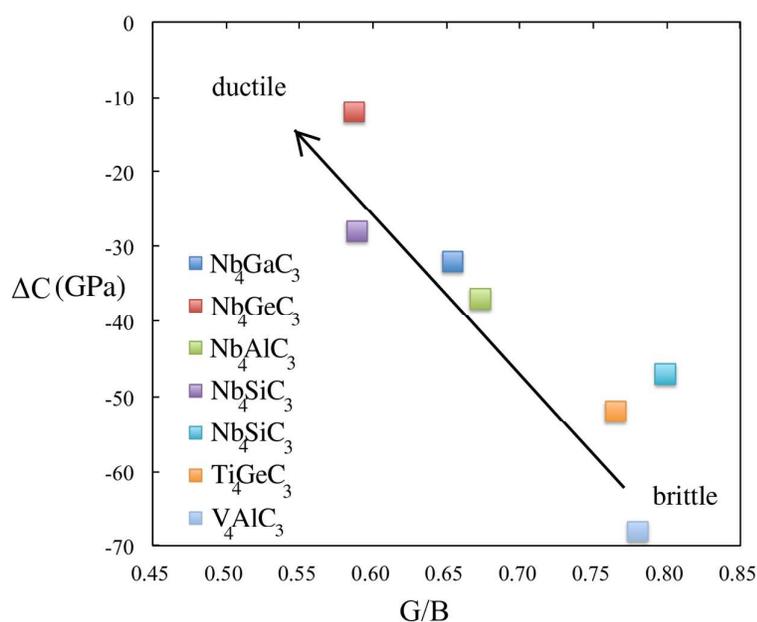

Fig. 4 (Color online) Map of brittleness and ductility trends in compounds evaluated in this work.

### 3.3. Electronic structure, Fermi surface and superconductive properties

In this section, the electronic structure of $Nb_4AlC_3$, $Nb_4SiC_3$, $Nb_4GaC_3$ and $Nb_4GeC_3$ are discussed and compared. The effects of different A elements in $M_4AlX_3$ are revealed by analyzing density of states (DOS) in Fig. 5 and Fermi surface in Fig. 6. It is clear that all of the studied $Nb_4AC_3$ phases present the metallic feature and the Nb-4*d* states mainly dominate at $E_f$. There are total two valence bands below $E_f$: One of them is below -10 eV, and the other is within the range from -10 eV to -0.5 eV. The latter is primarily composed of hybridization between Nb-4*d* and C-2*p* with energy range from -7.5 eV to -2.5 eV. A pseudo gap can be observed around -0.5 eV at the vicinity of $E_f$, and this gap separates the two valence bands. For the upper valence band, one of the strong $4d$-$2p$ (Nb-C) hybridization peaks is situated at -1 eV, and the other is situated at -2.5 eV. In the middle of the two $4d$-$2p$ (Nb-C) hybridization peaks, the evident $4d$-$p$ hybridization peak between Nb-A can also be found.

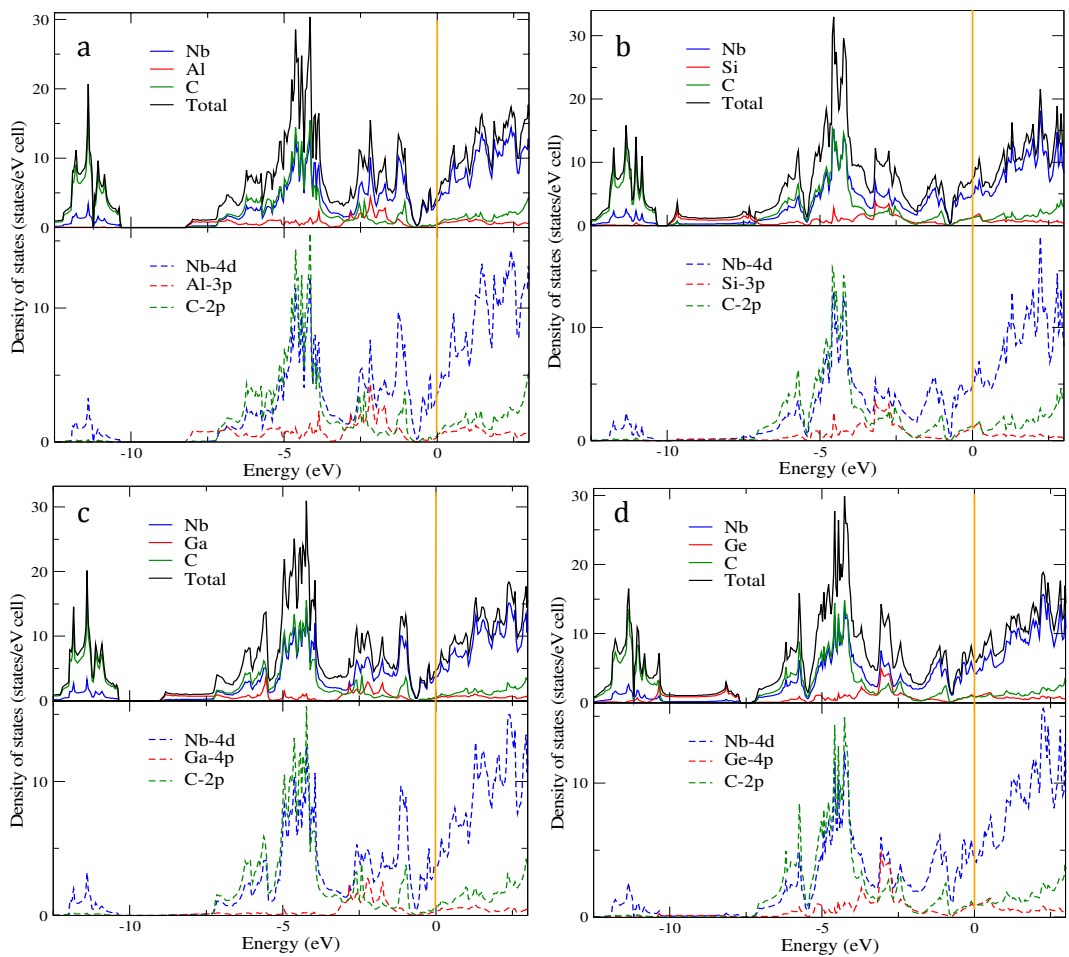

Fig. 5 Total, site-projected and partial DOSs of (a) $Nb_4AlC_3$, (b) $Nb_4SiC_3$ (c) $Nb_4GaC_3$ (d) $Nb_4GeC_3$. In each subfigure, the upper panel shows the site-projected and total DOS; and the lower panel shows the orbital-projected DOS of Nb-$4d$ states, A-$3p$/$4p$ states, and C-$2p$ states, respectively. The Fermi level ($E_f$) is set to be 0 eV as the orange horizontal line indicates.

More specifically, due to introducing different $p$ electrons from A element, the hybridization of Nb-$4d$ and C-$2p$ is mainly perturbed in the energy range from -2.5 eV to -0.5 eV, where the $3p$ ($4p$) states of A elements show overlapping with Nb-$4d$ and/or C-$2p$. The bond character of Nb-A is a mixture of ionicity and covalency. Judging from overlapping energy range, it is clear that the main character of Al (Ga) with Nb-$4d$ bond has more ionic feature. Si-$3p$ and Ge-$4p$ hybridizing with Nb-4*d* is evident, in the energy range of -3.0 to -2.5 eV, which indicates that covalent bonds between Si-Nb and Ge-Nb have been formed at that energy range. Therefore, the Si-Nb and Ge-Nb bonds have covalent part than those of Ga-Nb and Al-Nb. This can also be verified by accessing the excess electrons on Si and Ge in Table 4 (in the next section), where Si and Ge have obtained excess 0.79 and 0.88 electrons, and the values are greater than that of Al and Ga. Naturally, the smaller value of the excess electrons indicates a more ionic feature of the A-Nb bond. Some other difference are found in the lower valence band, in which the A states turn out to broaden much when one more electron fills in the *p* shell in the cases of Si and Ge. Concerning the effects of heavier *p* electron, we can easily identify that the extension of the tail of 3*p* states in Si heads towards the lower energy scale, while that of 4*p* states in Ge heads to upper energy scale. It should be emphasized that the tail composed of *p* states does not strongly hybridize with other states, which refers to the ionic feature of the *p* states in this range. For the phases having one more *p* electron, the 4*d*-3*p*(4*p*) hybridization can become much stronger reflected by the fact that the states split into more prevailing bonding and anti-bonding ones below -5 eV in Fig.5 (b) & (d). Overall, both of the filling one more *p* electron and having heavier *p* electron bring stronger Nb-A and Nb-C hybridization, which is considered as the underlying origin of bigger compressibility and better ductility.

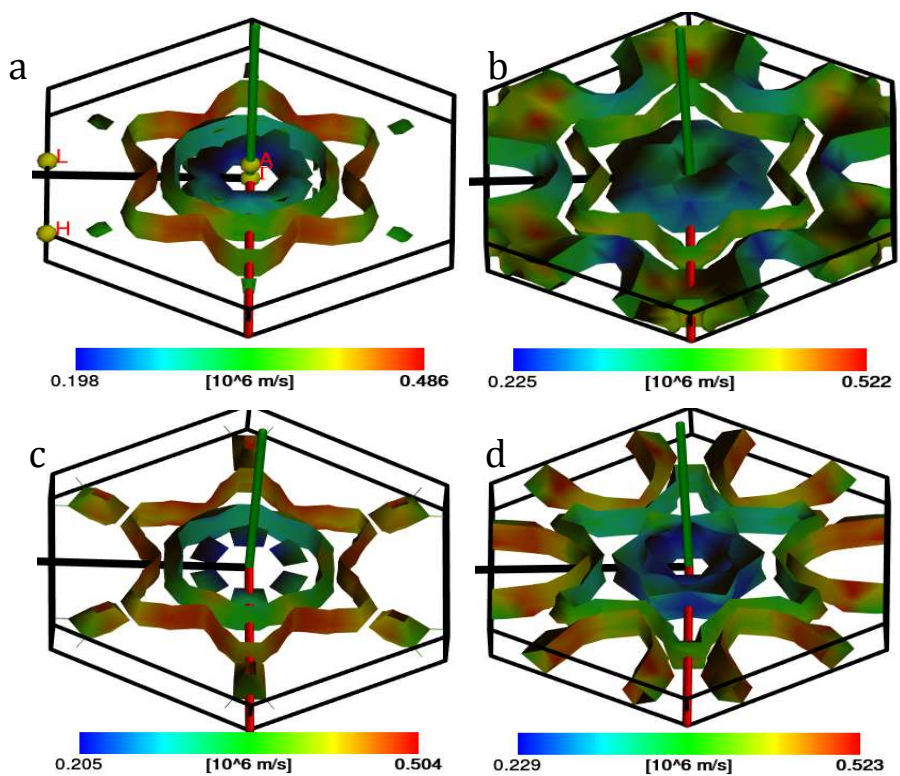

Fig. 6 Fermi surfaces of (a) $Nb_4AlC_3$ (b) $Nb_4SiC_3$ (c) $Nb_4GaC_3$ (d) $Nb_4GeC_3$ within the hexagonal cell (black) with high symmetry points in yellow, Γ, A, L, H in red (only shown in a). The red and black line represent Cartesian X,Y axis and the green line represents Cartesian Z axis. The Fermi velocity united by m/s is also shown.

As it is mentioned in the introduction, most of superconductors in MAX phases are Nb based, and it is figured out that the highest occupied states at $E_f$ ($N_{tot}(E_f)$) in these MAX phases superconductors reach 2.9 states/eV f.u. [28,29,30,31]. In the cases of the studied $Nb_4AC_3$ phases, the highest values of $N_{tot}(E_f)$, the states at $E_f$, is found in $Nb_4SiC_3$ with a value of 3.93 states/eV f.u. $N_{tot}(E_f)$ in $Nb_4SiC_3$ is much higher than that of the known MAX phases superconductors, which gives us a very strong signal of higher $T_c$. In Fig. 5 (b), it can be seen that the shift of $E_f$ is induced by the localization of hybridization between $4d$-$2p$ (Nb-Si). It can be implied that a higher occupation at $E_f$ also leads to a higher formation energy shown in the case of $Nb_4SiC_3$ shown in Table 1. However, it is known that Fermi surface (FS) is directly linked to superconductive properties, and in order to evaluate them, the calculation of FS of four compounds have been performed and results are shown in Fig. 6.

In Fig. 6, we can basically obtain the construction elements of FS, and our investigation extends from the simplest $Nb_4AlC_3$ to the most complicated $Nb_4SiC_3$. If $Nb_4AlC_3$ and $Nb_4GaC_3$ are shed into light, there are six pins near the BZ zone boundary in black, and a continuous two dimensional (2D) hexagonal sheet that are closed to the pins. In the BZ zone center, a complex has been discovered and this can be dependent on structures. Briefly, FS of $Nb_4AlC_3$ and $Nb_4GaC_3$ (Fig.6 a, c) are relatively simple and they also share some common features, such as the six separate pins outside, the neck constructed by 2D cylinder sheet. However, in the inner center, a connected 3D sheet is found in $Nb_4AlC_3$, and six 1D sheets in $Nb_4GaC_3$ are found in Fig. 6(c). On the right side of Fig. 6, the FS of $Nb_4SiC_3$ and $Nb_4GeC_3$ become more dispersive and more complicated. The sheets near the BZ boundary of $Nb_4SiC_3$, in comparison the separated spurs-like sheets in $Nb_4GeC_3$, become connected and grow to grooves shape, which is a typical nesting feature. The inner center also shows an outward fountain-like sheet near the zone center. The increased high $N_{tot}(E_f)$ and the nesting of FS in $Nb_4SiC_3$ propose a possible novel superconductor, and this may extend the list of superconductor of MAX phases.

**3.4. Optical properties and the unconventional anisotropic behavior**

The study of the optical functions of solids provides a better understanding of the electronic structure. The dielectric function, $\varepsilon(\omega) = \varepsilon_1(\omega) + i\varepsilon_2(\omega)$, describes the optical properties of a medium at full range of photon energies in a complex plane. The imaginary part of complex dielectric function can be derived from the momentum matrix elements between the occupied and the unoccupied states as shown below [46]:

$$\varepsilon_2(\omega) = \frac{2e^2\pi}{\Omega\varepsilon_0} \sum |\Psi_k^c | \mu \bullet r | \Psi_k^v |^2 \, \delta(E_k^c - E_k^v - E) \qquad \text{Equation (3)}$$

where $\Psi_k^c$ and $\Psi_k^v$ are the conduction and valence band wave functions at $k$, respectively, $\mu$ is the vector defining the polarization of the incident electric field, $c$ is the light frequency, and $e$ is the electronic charge. These compounds are crystallized in the $P6_3/mmc$ space group, which has two dominant and independent components of the dielectric tensors, which are $\varepsilon_2^{xx}(\omega)$ and $\varepsilon_2^{zz}(\omega)$. They respectively represent the components of the polarization parallel and perpendicular to the crystallographic c-axis. For the metallic materials there are two contribution to $\varepsilon(\omega)$, namely intra-band and inter-band transitions, written as $\varepsilon_2^\alpha(\omega) = \varepsilon_{2\,inter}^\alpha(\omega) + \varepsilon_{2\,intra}^\alpha(\omega)$. Owing to the metallic nature of the investigated MAX phases, we must include the Drude term (intra-band transitions) for $\varepsilon_2(\omega)$ [47]:

$$\varepsilon_{2\,intra}^\alpha(\omega) = \frac{\omega_p^2 \Gamma}{\omega(\omega^2 + \Gamma^2)} \qquad \text{Equation (4)}$$

where $\omega_p$ is the plasma frequency and $\Gamma = \hbar/\tau$, where $\tau$ is the mean free time between collisions (relaxation time), and $\alpha$ denotes the polarization orientation, $xx$, $yy$ or $zz$. The theoretical unscreened (Drude) plasma frequencies are derived from the Lindhard intra-band expression [48,49] with a damping of 0.15 eV. This value basically proportional to the squared root of Fermi velocity, and the results are presented in Table 3.

Table 3. The computed plasma frequencies for these four compounds in eV.

| Compounds | $Nb_4AlC_3$ | $Nb_4SiC_3$ | $Nb_4GaC_3$ | $Nb_4GeC_3$ |
|---|---|---|---|---|
| $\hbar\omega_p^{xx}$ (eV) | 3.2981 | 3.6879 | 3.5700 | 4.2541 |
| $\hbar\omega_p^{zz}$ (eV) | 1.9548 | 4.7533 | 2.4572 | 3.4839 |

In Table 3, it is clear that a strong anisotropy is present; and in most of the structures, the contribution from the basal plane is much stronger than that from out-of plane. For general layered structures, the contribution from the basal plane (*xx*, *yy*) is much bigger than that of *zz*. Nevertheless, the contribution from $\hbar\omega_p^{zz}$ is unconventionally bigger than that of *xx* component in $Nb_4SiC_3$, in contrast to the rest of structures and other known layered structures. As it is mentioned before, the Fermi velocity is contained in the plasma frequency that represents the contribution from different Fermi sheets in various orientations as shown in the following [50]:

$$(\omega_p^\alpha) = \frac{e^2}{\varepsilon_0} \int \frac{dk^3}{(2\pi)^3} \delta(\varepsilon_k - E_f)(v_k^\alpha)^2 \qquad \text{Equation (5)}$$

where $E_f$ stands for Fermi level and $v_k^\alpha$ is Fermi velocity, which is the origin of the anisotropic contribution from *k*. Then we move to the Fermi velocity tensor, the matrix element of *xx*, *yy*, *zz* components in the case of $Nb_4SiC_3$ are 4.69, 4.69 and 7.80. For the other cases, the *xx* and *yy* components are much bigger than *zz* component. This anomalous phenomenon reminds us of FS shown in Fig.6 (b), in which the center of the nesting sheets exhibits very high velocity (seen as red valley). As the Cartesian axis shows, the nesting is long Γ-A path, namely *c*-axis. Herein, the Fermi surface nesting with a high velocity along *c*-axis brings out unusual large plasma frequencies in $Nb_4SiC_3$, which ultimately results in the above unconventional optical properties.

From here, the real and imaginary parts of dielectric function of all these four structures are discussed and plotted in Fig. 7, 8 respectively. When the imaginary dielectric function is produced, the real part of the dielectric function is calculated directly using Kramers–Kronig relations [51]. Owing to the strong anisotropic lattice feature, the real and imaginary parts of dielectric function are both shown in different polarization vector <100> (in plane) and <001> (out-of plane) in order to obtain the comprehensive optical properties. Since $Ti_4AlN_3$ was reported to be used as a coating on spacecraft to avoid solar heating [52], the reflectivity will be shown as well in Fig.9. The focus will be on the reflectance in the infrared and closed infrared regions. With the inclusion of Drude term correction, the calculated optical properties will give a more precise description of $Nb_4AC_3$ structures. The relation between anisotropic behaviors, i.e. the lattice and the Fermi surface to the dielectric function is of interest to unravel.

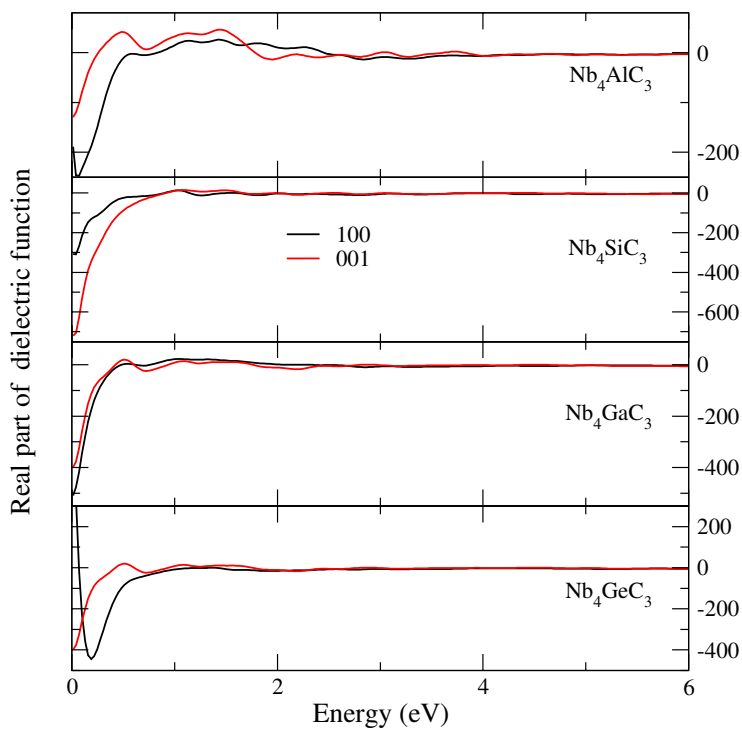

Fig.7. Real part of dielectric function of $Nb_4AlC_3$, $Nb_4SiC_3$, $Nb_4GaC_3$ and $Nb_4GeC_3$ from the top panel to the bottom. The black cures stands for the dielectric function along <100> direction and red curves for <001> direction. The blue rectangle highlights the low energy scale.

Despite a few different points exhibit in the real part of dielectric function in Fig. 7, the common feature of the spectra start from a negative value to 0 eV, and then to a positive value around 1 eV, approaching to be constant about 2 eV. At low energy range, intra-band transition dominates and has a negative contribution. As the photon energy increases the intensity of the spectrum increase to zero suggesting the appearance of plasma. Furthermore, the spectrum of <100> is below that of <001> direction as a result of greater plasma frequency in <100> direction and a more negative contribution to the spectrum, which demonstrates the anisotropic of hexagonal lattice. Nevertheless, $Nb_4SiC_3$ shows an opposite tendency compared with the others, the spectrum of <100> above <001>, and this can be driven by the large contribution from Z direction of intra-band transition (lower starting point at the static limit). Note that the Drude correction mainly lays in the low energy scale, and this phenomenon occurred in $Nb_4SiC_3$ coincides with the observation of large contribution from the *zz* component of plasma frequency in $Nb_4SiC_3$. One more interesting observable is that the high peak at the static limit is found in $Nb_4GeC_3$, although a negative contribution on the dielectric function from the intra-band transition is introduced. The positive of value may be due to large positive *d-d* inter-band transitions, which cancels the negative inter-band transition. When photon energy increases, the both intra- and inter- band transitions contribute negatively to the spectrum. Therefore, the spectrum decreases rapidly to a large negative value. The big negative values of $\varepsilon_1(\omega)$ at low energy scale in $Nb_4AlC_3$, $Nb_4SiC_3$ and $Nb_4GaC_3$ indicate that they show typical Drude-like behavior, and the behavior of the two transitions along <100> direction in $Nb_4AlC_3$ and $Nb_4GeC_3$ are very similar, both of which have valleys (turning points) close to the static limit.

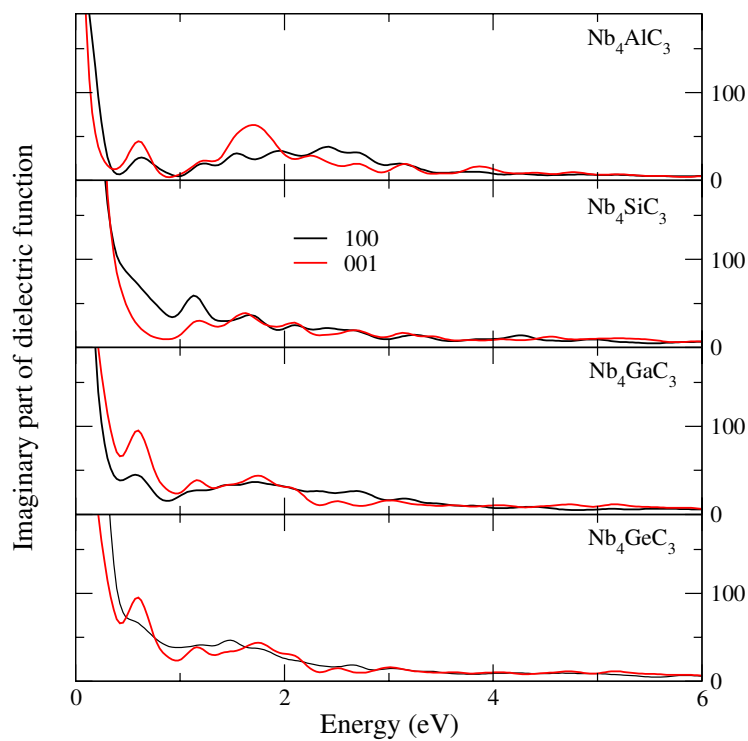

Fig.8 Imaginary part of dielectric function of $Nb_4AlC_3$ (a), $Nb_4SiC_3$ (b), $Nb_4GaC_3$ (c) and $Nb_4GeC_3$ (d). The black cures stands for the dielectric function along <100> direction and red curves for <001> direction. The blue and green rectangles highlight the two energy regions.

Followed by the real part of dielectric function, the imaginary part of dielectric function of these four materials is shown in Fig. 8. When the photon energy is smaller than 0.5 eV, the imaginary part is mainly contributed by intra-band transition, and at 0.6 eV the first strong *d-d* inter-band transition peak is found after the region governed by Drude correction. Note that the discussion on the DOS (relative shift of $E_f$ to posses higher occupation) and the unusual contribution from Z direction makes $Nb_4SiC_3$ exhibit peculiar transitions shown in Fig.8. However, unlike the other three materials, the first strong transition does not occur at 0.6 eV but is pushed further to 1.2 eV for both polarization orientations in Nb$_4$SiC$_3$. It might be due to different features appearing in the PDOS of Nb $4d$ states around Fermi level inducing different optical transitions. Furthermore, this transition along <100> direction is stronger than that along <001> direction, which shows in the opposite way from the others. For the secondary strong transition located around 1.7 eV, contributions from the two distinctive directions are comparable in Nb$_4$SiC$_3$. While for the others, the contribution from <001> is more pronounced, which can be reflected by plasma frequencies shown in Table 3. In fact, these two evident peaks can link to the strong hybridization near $E_f$, between Nb-A shown in Fig. 5.

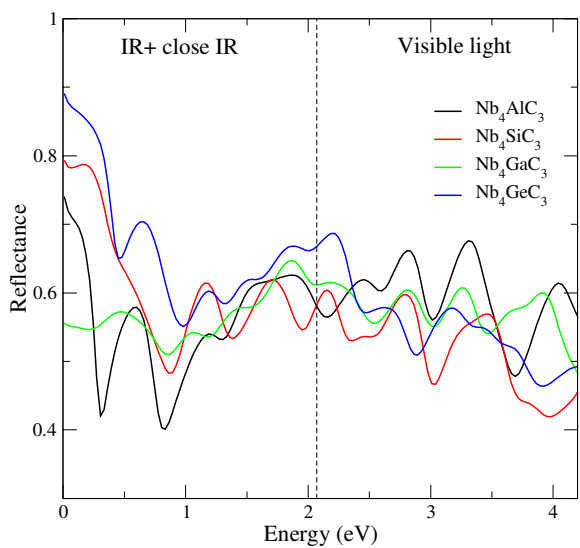

Fig.9 The reflectance of Nb$_4$AlC$_3$ (black curves), Nb$_4$SiC$_3$ (red curves), Nb$_4$GaC$_3$ (green curves) and Nb$_4$GeC$_3$ (violet curves). The dotted vertical line divides the energy range into infrared and close infrared region (IR + close IR), and visible light region.

As shown in Fig. 9, these four materials reach the highest reflectance near the static limit, and around 0.5 eV the entire reflectance drop down, which is induced by Drude model. In fact, the dip is so-called plasma edge, a screened plasma frequency by the infinite dielectric constant $\varepsilon_\infty$ ($\hbar w_p^{scr} \approx \frac{\hbar w_p}{\sqrt{\varepsilon_\infty}}$). In the sense of variation of $p$ electrons, the number of electrons in $p$ shell in A elements determines the reflectance spectra. It is known that NbC, this ternary carbide can make the efficient improvement to the reflectance to the infrared region (IR) as radiant barrier coatings. This implies that solar heating can be reduced the infrared emittance, and the $Nb_4GeC_3$ surface will therefore be moderate in strong sunlight. The reflectance of $Nb_4GeC_3$ in the IR region is higher in contrast to the other three structures (the average value is above 0.7), and this high reflectance in $Nb_4GeC_3$ even shows better performance than $Ti_4AlN_3$[52]. The combination of good reflectance and excellent mechanical strengths under high temperature, make $Nb_4GeC_3$ perhaps have a potential application in solar heating coatings. While the reflectance of $Nb_4GaC_3$ is non-sensitive over the photon energy range as we explored, it can be due to the characteristic of chemical bonding.

If the valence electrons participate in strong bonding, e.g. covalent bond, the electrons are about to delocalized further leading to polarization, which will enhance the reflectance. In the electronic structure in Fig. 5, at the $E_f$, the main contribution is from Nb-4d states as well as hybridization of *pp* (A-C) hybridization. *(Answer: Yes, there are two main contributions)* Hereby the deduction moves to the effects of A elements again. In this sense, the excess number of electrons are plausible to investigate and listed in table 4.

Table 4.The computed excess electron on of A element (per atom).

| Compounds | $Nb_4AlC_3$ | $Nb_4SiC_3$ | $Nb_4GaC_3$ | $Nb_4GeC_3$ |
|---|---|---|---|---|
| Excess electron on Nb-4f site | -1.77 | -1.77 | -1.79 | -1.79 |
| Excess electron on Nb-4e site | -1.22 | -1.34 | -1.13 | -1.40 |
| Excess electron on A-2c site | 0.53 | 0.79 | 0.36 | 0.88 |
| Excess electron on C-2a site | 1.72 | 1.73 | 1.76 | 1.75 |
| Excess electron on C-4f site | 1.87 | 1.85 | 1.87 | 1.88 |

The number of excess electron can be regarded as the probability of charge transfer. It can be demonstrated that between the two type atoms of C and Nb-4f, the number of excess electron remain the same level, but only charge transfer between Nb (4e site)-A varies in an observable amount. The synchronous variation of charge transfer between Nb (4e site)-A interactions is found to be strongly perturbed by A elements. In general, the bigger of the charge transfer, the higher concentration of ionic feature. The strong transition signal generally results from the strong hybridization, and is mostly covalent bonding shown in Eq. 3. It can be seen that $Nb_4GeC_3$ shows the stronger transition at the static limit, which is confirmed by the biggest number of excess electrons in Table 4. It is therefore straightforward to understand why the reflectance in $Nb_4GeC_3$ is rather weak. It is because that Nb (4e site)-Ga forms a bond with more likely ionic feature, which can be reflected by the rather small number of excess number of electrons on Nb (4e site) and Ga site. Overall, the Drude type behaviors are identified, and the unusual anisotropic optical properties along c-axis in $Nb_4SiC_3$ are discovered. In the view of applications, $Nb_4GeC_3$ is a potential candidate as a solar-heating barrier material. It can be concluded that the perturbation of A elements on optical properties is of importance and could be used to control the performance.

## 4. Conclusion

In the present work, three new MAX phases $Nb_4SiC_3$, $Nb_4GaC_3$ and $Nb_4GeC_3$ have been predicted to be mechanically stable. A general principle of A elements on various properties have been proposed. All three phases energetically favored to crystallize in the α-type crystal structure of $M_4AX_3$, and the negative formation energy further verifies its thermodynamical stability. All the relaxed structural information has been reported, and the typical anisotropic lattices are present here. In the diffraction patterns, a few new peaks other than the ones in $Nb_4AlC_3$ emerge, which can be regarded as the fingerprints to distinguish different structures.

They all show good compressibility reflected by large bulk modulus (above 200 GPa), and the excellent stiffness, *etc*. It can be implied that the more and heavier *p* electrons in A elements can enhance the ductility and the anisotropic elasticity, and this tendency can give essential guideline to industrial application of MAX phases on the mechanical usage.

The metallic feature and typical $4d$-$2p$ hybridization between Nb-C are shown in the density of states. The more and heavier *p* electrons will drives the DOS picture downward and the Nb-A interaction makes it more localized, which is the origin of shift of $N(E_f)$. These changes in the electronic structure due to the variation of A elements also corresponds to the behavior of several mechanical properties. It is found out that the high population at $E_f$ as well as the nesting of Fermi surface in $Nb_4SiC_3$ releases a very strong signal of superconductor. In turn, the extremely high Fermi velocity perpendicular to the basal plane contributing to large plasma frequency considered in Drude model induces the unconventional behavior in $Nb_4SiC_3$. In other words, the large contribution from out-of plane by introducing the Drude correction breaks the typical anisotropic behavior in other $Nb_4AC_3$, and the relevant optical properties turn to be more *isotropic* in the *anisotropic* hexagonal lattice.

The unconventional optical properties in $Nb_4SiC_3$ are first reported in MAX phases, which may bring out more interesting behaviors, such as transport properties. It is also showed that $Nb_4GeC_3$ could serve as a coating material to avoid solar heating, so that $Nb_4GeC_3$ is a candidate material for coatings in future space missions to Mercury. We hope that the theoretical predictions will inspire experimental investigation on multifunctional $Nb_4SiC_3$ and $Nb_4GeC_3$.

## 5. Acklwdgement

Fu Yu-dong is grateful to the Fundamental Research Funds for the Central Universities of P.R.C CHN (HEUCF20151015). Weiwei Sun is grateful to the High Performance Computing Center North (HPC2N) in Sweden for computational time.